\documentclass[conference]{IEEEtran}

\usepackage{amsmath}
\usepackage{amssymb}
\usepackage[dvips]{graphicx}
\usepackage{epsfig}

\newcommand{\x}{\mathbf{x}}
\newcommand{\y}{\mathbf{y}}

\newcommand{\z}{\mathbf{z}}

\newcommand{\n}{\mathbf{n}}

\newcommand{\I}{\mathbf{I}}

\newcommand{\dd}{\dagger}

\newcommand{\he}{\hat{h}}
\newcommand{\hn}{\tilde{h}}
\newcommand{\CN}{\mathcal{CN}}

\newcommand{\figsize}{0.35}
\newcommand{\ffigsize}{0.4}

\newcommand{\tsnr}{{\text{\footnotesize{SNR}}}}

\newtheorem{lemma:bitenergylow}{Lemma}
\newtheorem{lemma:bitenergyhigh}[lemma:bitenergylow]{Lemma}

\newtheorem{prop:asympcap}{Theorem}
\newtheorem{prop:flashminbitenergy}[prop:asympcap]{Proposition}
\newtheorem{prop:flashbitenergy}[prop:asympcap]{Theorem}
\newtheorem{prop:pasympcap}[prop:asympcap]{Theorem}

\begin{document}

\title{An Energy Efficiency Perspective on Training for Fading Channels}



%
\author{\authorblockN{Mustafa Cenk Gursoy}
\authorblockA{Department of Electrical Engineering\\
University of Nebraska-Lincoln, Lincoln, NE 68588\\ Email:
gursoy@engr.unl.edu}}


\maketitle

\begin{abstract}\footnote{This work was supported in part by the NSF CAREER Grant
CCF-0546384.} In this paper, the bit energy requirements of
training-based transmission over block Rayleigh fading channels are
studied. Pilot signals are employed to obtain the minimum
mean-square-error (MMSE) estimate of the channel fading
coefficients. Energy efficiency is analyzed in the worst case
scenario where the channel estimate is assumed to be perfect and the
error in the estimate is considered as another source of additive
Gaussian noise. It is shown that bit energy requirement grows
without bound as the $\tsnr$ goes to zero, and the minimum bit
energy is achieved at a nonzero $\tsnr$ value below which one should
not operate. The effect of the block length on both the minimum bit
energy and the $\tsnr$ value at which the minimum is achieved is
investigated. Flash training schemes are analyzed and shown to
improve the energy efficiency in the low-$\tsnr$ regime. Energy
efficiency analysis is also carried out when peak power constraints
are imposed on pilot signals.
\end{abstract}

\section{Introduction}

One of the challenges of wireless systems is to establish
communication under time-varying channel conditions experienced due
to mobility and changing environment. One technique employed in
practical systems to cope with this challenge is to periodically
send training signals to estimate the channel. Often, the channel
estimate is considered as perfect, and transmission and reception is
designed for a known channel. Due to its practical significance,
training schemes has been studied extensively. Tong \emph{et al.} in
\cite{Tong} present an overview of pilot-assisted wireless
transmissions and discuss design issues from both
information-theoretic and signal processing perspectives. The
information-theoretic approach considers the optimization of
training parameters to maximize the achievable data rates. For
instance, Hassibi and Hochwald \cite{Hassibi} optimized the training
data, power, and duration in multiple-antenna wireless links by
maximizing a lower bound on the channel capacity. The general theme
in information-theoretic approaches (see e.g., \cite{Tong} and
references therein) is that training-based transmission schemes are
close to being optimal at high signal-to-noise ($\tsnr$) values but
highly suboptimal in the low-$\tsnr$ regime due to poor channel
estimates.

Another important concern in wireless communications is the
efficient use of limited energy resources. In systems where energy
is at a premium, minimizing the energy cost per unit transmitted
information will improve the efficiency. Hence, the energy required
to reliably send one bit is a metric that can be adopted to measure
the performance. Generally, energy-per-bit requirement is minimized,
and hence the energy efficiency is maximized, if the system operates
in the low-$\tsnr$ regime \cite{Verdu}, \cite{Gursoy-part2}. Since
training-based schemes perform poorly at low $\tsnr$s especially if
the channel estimate is presumed to be perfect, this immediately
calls into question the energy-efficiency of pilot-assisted systems.
With this motivation, we present in this paper an energy-efficiency
perspective on pilot-assisted wireless transmission schemes and seek
to answer the question of how low should the $\tsnr$ be so that the
energy efficiency is compromised.

\section{Channel Model} \label{sec:channelmodel}

We consider Rayleigh block-fading channels where the input-output
relationship within a block of $m$ symbols is given by
\vspace{-.2cm}
\begin{gather}\label{eq:model}
\y = h \x + \n 
\end{gather}
where $h \sim \mathcal{CN}(0,\gamma^2)$ \footnote{$\x \sim
\mathcal{CN}(\mathbf{d}, \mathbf{\Sigma})$ is used to denote that
$\x$ is a complex Gaussian random vector with mean $E\{\x\} =
\mathbf{d}$ and covariance $E\{(\x - \mathbf{d})(\x -
\mathbf{d})^\dd\} = \mathbf{\Sigma}$} is a zero-mean circularly
symmetric complex Gaussian random variable with variance $E\{|h|^2\}
= \gamma^2$, and $\n$ is a zero-mean, $m$ complex-dimensional
Gaussian random vector\footnote{Note that in the channel model
(\ref{eq:model}), $\y$, $\x$, and $\n$ are column vectors.} with
covariance matrix $E\{\n \n^\dd\} = N_0 \I$. $\x$ and $\y$ are the
$m$ complex-dimensional channel input and output vectors
respectively. The input is subject to an average power constraint
\vspace{-.2cm}
\begin{gather} \label{eq:avgpower}
E\{\|\x\|^2\} \le mP. 
\end{gather}
It is assumed that the fading coefficients stay constant for a block
of $m$ symbols and have independent realizations for each block. It
is further assumed that neither the transmitter nor the receiver has
prior knowledge of the realizations of the fading coefficients.

\section{Training-Based Transmission and Reception}
\label{sec:trainingscheme}

We assume that pilot symbols are employed in the system to
facilitate channel estimation at the receiver. Hence, the system
operates in two phases, namely training and data transmission. In
the training phase, pilot symbols known at the receiver are sent
from the transmitter and the received signal is
\begin{gather}
\y_t = h \x_t + \n_t
\end{gather}
where $\y_t$, $\x_t$, and $\n_t$ are $l$-dimensional vectors
signifying the fact that $l$ out of $m$ input symbols are devoted to
training. It is assumed that the receiver employs minimum
mean-square error (MMSE) estimation to obtain the estimate
\begin{gather}
\he = E\{h | \y_t\} = \frac{\gamma^2}{\gamma^2 \|\x_t\|^2 + N_0}
\x_t^\dd \y_t.
\end{gather}
With this estimate, the fading coefficient can now be expressed as
\begin{gather}
h = \he + \hn
\end{gather}
where
\begin{align}
\he \sim \CN \left( 0, \frac{\gamma^4 \|\x_t\|^2}{\gamma^2
\|\x_t\|^2 + N_0}\right) \intertext{and} \hn \sim \CN \left( 0,
\frac{\gamma^2 N_0 }{\gamma^2 \|\x_t\|^2 + N_0}\right).
\end{align}
Following the training phase, the transmitter sends the
($m-l$)-dimensional data vector $\x_d$ and the receiver equipped
with the knowledge of the channel estimate operates on the received
signal
\begin{gather} \label{eq:dataphasemodel}
\y_d = \he \x_d + \hn \x_d + \n_d
\end{gather}
to recover the transmitted information.

\section{Energy Efficiency in the Worst Case Scenario}

\subsection{Average Power Limited Case}

Our overall goal is to identify the bit energy values that can be
attained with optimized training parameters such as the power and
duration of pilot symbols. The least amount of energy required to
send one information bit reliably is given by\footnote{Note that
$\frac{E_b}{N_0}$ is the bit energy normalized by the noise power
spectral level $N_0$.}
\begin{gather}
\frac{E_b}{N_0} = \frac{\tsnr}{C(\tsnr)}
\end{gather}
where $C(\tsnr)$ is the channel capacity in bits/symbol. In general,
it is difficult to obtain a closed-form expression for the capacity
of the channel (\ref{eq:dataphasemodel}). Therefore, we consider a
lower bound on the channel capacity by assuming that
\begin{gather}
\z = \hn \x_d + \n_d
\end{gather}
is a Gaussian noise vector that has a covariance of
\begin{gather}
E\{\z \z^\dd\} =\sigma_{\hn}^2 E\{\x_d \x_d^\dd\} + N_0 \I,
\end{gather}
and is uncorrelated with the input signal $\x_d$. With this
assumption, the channel model becomes
\begin{gather} \label{eq:worstcasemodel}
\y_d = \he \x_d + \z.
\end{gather}
This model is called the worst-case scenario since the channel
estimate is assumed to be perfect, and the noise is modeled as
Gaussian, which presents the worst case \cite{Hassibi}. The capacity
of the channel in (\ref{eq:worstcasemodel}), which acts as a lower
bound on the capacity of the channel in (\ref{eq:dataphasemodel}),
is achieved by a Gaussian input with
$$E\{\x_d \x_d^\dd\} = \frac{(1-\delta^*)mP}{m-1} \I$$ where $\delta^*$ is
the optimal fraction of the power allocated to the pilot symbol,
i.e., $|x_t|^2 = \delta^* mP$. The optimal value is given by
\begin{gather}
\delta^* = \sqrt{\eta(\eta+1)} - \eta
\end{gather}
where
\begin{gather} \label{eq:eta+snr}
\eta = \frac{m \, \tsnr +(m-1)}{m(m-2) \tsnr} \quad \text{and} \quad
\tsnr = \frac{\gamma^2 P}{N_0}.
\end{gather}
Note that $\tsnr$ in (\ref{eq:eta+snr}) is the received
signal-to-noise ratio. In the average power limited case, sending a
single pilot is optimal because instead of increasing the number of
pilot symbols, a single pilot with higher power can be used and a
decrease in the duration of the data transmission can be avoided.
Hence, the optimal $\x_d$ is an ($m-1$)-dimensional Gaussian vector.
Since the above results are indeed special cases of those in
\cite{Hassibi}, the details are omitted. The resulting capacity
expression\footnote{Unless specified otherwise, all logarithms are
to the base $e$.} is
\begin{align}
C_L(\tsnr) &= \frac{m\!-\!\!1}{m} E_w \left\{ \log \left( 1 +\!
\frac{\phi(\tsnr)\tsnr^2}{\psi(\tsnr)\tsnr + (m\!-\!\!1)}
|w|^2\right) \right\} \nonumber
\\
&= \frac{m\!-\!\!1}{m} E_w \left\{ \log \left( 1 + f(\tsnr)
|w|^2\right) \right\} \text{ nats/symbol} \label{eq:worstcap}
\end{align}
where \vspace{-.4cm}
\begin{gather}
\phi(\tsnr) = \delta^*(1-\delta^*)m^2,
\\
\psi(\tsnr) = (1 + (m-2)\delta^*)m,
\end{gather}
and $w \sim \CN(0,1)$. Note also that the expectation in
(\ref{eq:worstcap}) is with respect to the random variable $w$. The
bit energy values in this setting are given by
\begin{gather} \label{eq:worstengy}
\frac{E_{b,U}}{N_0} = \frac{\tsnr}{C_L(\tsnr)} \log2.
\end{gather}
$\frac{E_{b,U}}{N_0}$ provides the least amount of normalized bit
energy values in the worst-case scenario and also serves as an upper
bound on the achievable bit energy levels of channel
(\ref{eq:dataphasemodel}). It is shown in \cite{Lapidoth Shamai}
that if the channel estimate is assumed to be perfect, and Gaussian
codebooks designed for known channels are used, and scaled nearest
neighbor decoding is employed at the receiver, then the generalized
mutual information has an expression similar to (\ref{eq:worstcap})
(see \cite[Corollary 3.0.1]{Lapidoth Shamai}). Hence
$\frac{E_{b,U}}{N_0}$ also gives a good indication of the energy
requirements of a system operating in this fashion. The next result
provides the asymptotic behavior of the bit energy as $\tsnr$
decreases to zero.

\begin{lemma:bitenergylow} \label{lemma:bitenergylow}
The normalized bit energy (\ref{eq:worstengy}) grows without bound
as the signal-to-noise ratio decreases to zero, i.e.,
\begin{gather}
\left.\frac{E_{b,U}}{N_0}\right|_{C_L = 0} = \lim_{\tsnr \to 0}
\frac{\tsnr}{C_L(\tsnr)}\log2 = \frac{\log2}{\dot{C}_L(0)} = \infty.
\end{gather}
\end{lemma:bitenergylow}
\vspace{.5cm} \emph{Proof}: In the low SNR regime, we have
\begin{align}
C_L(\tsnr) &= \frac{m-1}{m} \left( f(\tsnr) E\{|w|^2\} +
o(f(\tsnr))\right)
\\
&= \frac{m-1}{m} \left( f(\tsnr) + o(f(\tsnr))\right).
\end{align}
As $\tsnr \to 0$, $\delta^* \to 1/2$, and hence $\phi(\tsnr) \to m^2
/4$ and $\psi(\tsnr) \to m + m(m-2)/2$. Therefore, it can easily be
seen that \vspace{-.2cm}
\begin{gather}
f(\tsnr) = \frac{m^2}{4(m-1)} \tsnr^2 + o(\tsnr^2)
\end{gather}
from which we have $\dot{C}_L(0) = 0$. \hfill $\square$

The fact that $C_L$ decreases as $\tsnr^2$ as $\tsnr$ goes to zero
has already been pointed out in \cite{Hassibi}. Lemma
\ref{lemma:bitenergylow} shows the impact of this behavior on the
energy-per-bit, and indicates that it is extremely
energy-inefficient to operate at very low $\tsnr$ values.
We further conclude that in a training-based scheme where the
channel estimate is assumed to be perfect, the minimum energy per
bit is achieved at a finite and nonzero $\tsnr$ value. This most
energy-efficient operating point can be obtained by numerical
analysis. We can easily compute $C_L(\tsnr)$ in (\ref{eq:worstcap}),
and hence the bit energy values, using the Gauss-Laguerre quadrature
integration technique.
\begin{figure}
\begin{center}
\includegraphics[width = \ffigsize\textwidth]{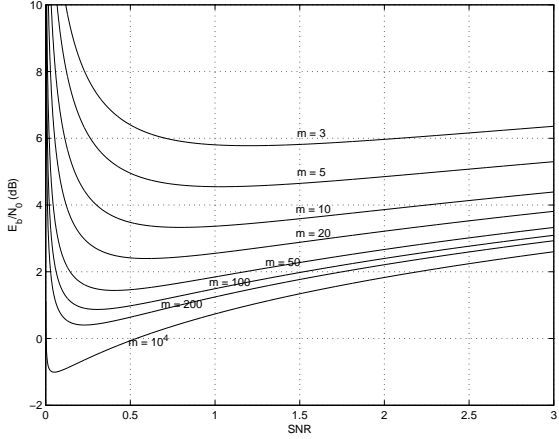}
\caption{Energy per bit $E_{b,U}/N_0$ vs. $\tsnr$ in the worst-case
scenario} \label{fig:trainingbitenergy}
\end{center}
\end{figure}

Figure \ref{fig:trainingbitenergy} plots the normalized bit energy
curves as a function of $\tsnr$ for block lengths of $m = 3, 5, 10,
20, 50, 100, 200, 10^4$. As predicted, for each block length value,
the minimum bit energy is achieved at nonzero $\tsnr$, and the bit
energy requirement increases as $\tsnr \to 0$. 
It is been noted in \cite{Hassibi} that training-based
schemes, which assume the channel estimate to be perfect, perform
poorly at very low $\tsnr$ values, and the exact transition point
below which one should not operate in this fashion is deemed as not
clear. Here, we propose the $\tsnr$ level at which the minimum bit
energy is achieved as a transition point since operating below this
point results in higher bit energy requirements.
Another observation from Fig. \ref{fig:trainingbitenergy} is that
the minimum bit energy decreases with increasing $m$ and is achieved
at a lower $\tsnr$ value. The following result sheds a light on the
asymptotic behavior of the capacity as $m \to \infty$.
\begin{prop:asympcap} \label{prop:asympcap}
As the block length $m$ increases, $C_L$ approaches to the capacity
of the perfectly known channel, i.e.,\vspace{-.1cm}
\begin{gather}
\lim_{m \to \infty} C_L(\tsnr) = E_w\{ \log(1 + \tsnr |w|^2)\}.
\end{gather}
Moreover, define $\chi = 1/m$. Then
\begin{gather}
\left.\frac{dC_{L}(\tsnr)}{d\chi}\right|_{\chi = 0} = -\infty.
\label{eq:derivative}
\end{gather}
\end{prop:asympcap}

\vspace{0cm} \emph{Proof}: We have
\begin{align}
\lim_{m \to \infty} C_L(\tsnr) &= \lim_{m \to \infty} E_w \left\{
\log \left( 1 + f(\tsnr) |w|^2\right) \right\} \label{eq:asymp1}
\\
&= E_w \left\{ \lim_{m \to \infty} \log \left( 1 + f(\tsnr)
|w|^2\right) \right\} \label{eq:asymp2}
\\
&= E_w \left\{ \log \left( 1 + \tsnr |w|^2 \right) \right\}.
\label{eq:asymp4}
\end{align}
(\ref{eq:asymp2}) holds due to integrable upper bound $\left| \log(1
\!\!+\!\! f(\tsnr)|w|^2)\right| \le 3\tsnr |w|^2$ and the Dominated
Convergence Theorem. (\ref{eq:derivative}) follows again from the
application of the Dominated Convergence Theorem and the fact that
the derivative of $f(\tsnr)$ with respect to $\chi = 1/m$ at $\chi =
0$ is $-\infty$.
\hfill $\square$

\begin{figure}
\begin{center}
\includegraphics[width = \ffigsize\textwidth]{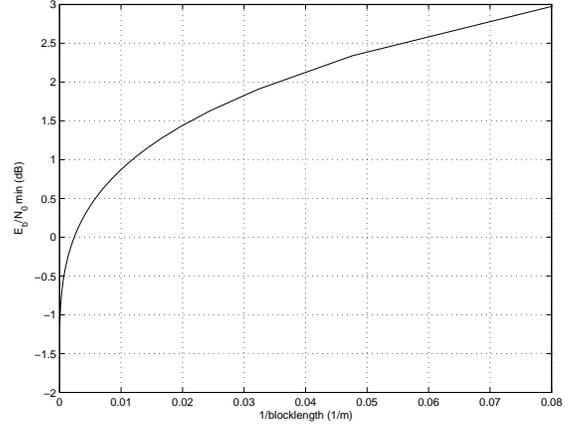}
\caption{Minimum energy per bit $\frac{E_{b,U}}{N_0}_{min}$ vs.
$\frac{1}{m}$ in the worst-case scenario} \label{fig:minbit_vs_m}
\end{center}
\end{figure}

The first part of Theorem \ref{prop:asympcap} is not surprising and
is expected because reference \cite{Marzetta} has already shown that
as the block length grows, the perfect knowledge capacity is
achieved even if no channel estimation is performed.
This result agrees with our observation in Fig.
\ref{fig:trainingbitenergy} that $-1.59$ dB is approached at lower
SNR values as $m$ increases. 
However, the rate of approach is very slow in terms of the block
size, as proven in the second part of Theorem \ref{prop:asympcap}
and evidenced in Fig. \ref{fig:minbit_vs_m}. Due to the infinite
slope\footnote{Note that Theorem \ref{prop:asympcap} implies that
the slope of $\frac{\tsnr}{C_L(\tsnr)}$ at $\chi = \frac{1}{m} = 0$
is $\infty$.} observed in the figure, approaching $-1.59$dB is very
demanding in block length.

\subsection{Flash Training and Transmission}

One approach to improve the energy efficiency in the low $\tsnr$
regime is to increase the peak power of the transmitted signals.
This can be achieved by transmitting $\nu$ fraction of the time with
power $P/\nu$. Note that training also needs to be performed only
$\nu$ fraction of the time. This type of training and communication,
called flash transmission scheme, is analyzed in \cite{Zheng} where
it is shown that the minimum bit energy of $-1.59$dB can be achieved
if the block length $m$ increases at a certain rate as $\tsnr$
decreases. In the setting we consider, flash transmission scheme
achieves the following rate:
\begin{gather}
C_{fL}(\tsnr, \nu) = \nu(\tsnr) C_L\left(
\frac{\tsnr}{\nu(\tsnr)}\right)
\end{gather}
where $0 < \nu(\cdot) \le 1$ is the duty cycle which in general is a
function of the $\tsnr$. First, we show that flash transmission
using peaky Gaussian signals does not improve the minimum bit
energy.
\begin{prop:flashminbitenergy} \label{prop:flashminbitenergy}
For any duty cycle function $\nu(\cdot)$,
\begin{gather}
\inf_\tsnr \frac{\tsnr}{C_{fL}(\tsnr,\nu)} \ge \inf_\tsnr
\frac{\tsnr}{C_{L}(\tsnr)}.
\end{gather}
\end{prop:flashminbitenergy}

\vspace{.5cm} \emph{Proof}: Note that for any $\tsnr$ and
$\nu(\tsnr)$,
\begin{align}
\!\!\!\!\!\!\!\!\!\!\!\!\!\!\frac{\tsnr}{C_{fL}(\tsnr,\nu)} =
\frac{\frac{\tsnr}{\nu(\tsnr)}}{C_L\left(
\frac{\tsnr}{\nu(\tsnr)}\right)}
=\frac{\tilde{\tsnr}}{C_L(\tilde{\tsnr})} \ge \inf_\tsnr
\frac{\tsnr}{C_{L}(\tsnr)} \label{eq:flashineq}
\end{align}
where $\tilde{\tsnr}$ is defined as the new $\tsnr$ level. Since the
inequality in (\ref{eq:flashineq}) holds for any $\tsnr$ and
$\nu(\cdot)$, it also holds for the infimum of the left-hand side of
(\ref{eq:flashineq}), and hence the result follows. \hfill $\square$

We classify the duty cycle function into three categories:
\begin{enumerate}
\item $\nu(\cdot)$ that satisfies $\lim_{\tsnr \to 0} \frac{\tsnr}{\nu(\tsnr)} = 0$
\item $\nu(\cdot)$ that satisfies $\lim_{\tsnr \to 0} \frac{\tsnr}{\nu(\tsnr)} = \infty$
\item $\nu(\cdot)$ that satisfies $\lim_{\tsnr \to 0}
\frac{\tsnr}{\nu(\tsnr)} = a$ for some constant $a > 0$.
\end{enumerate}
Next, we analyze the performance of each category of duty cycle
functions in the low-$\tsnr$ regime.

\begin{prop:flashbitenergy} \label{prop:flashbitenergy}
If $\nu(\cdot)$ is chosen from either Category 1 or 2,
\begin{gather}
\left.\frac{E_{b,U}}{N_0}\right|_{C_{fL} = 0} = \lim_{\tsnr \to 0}
\frac{\tsnr}{C_{fL}(\tsnr, \nu)}\log2 = \infty.
\end{gather}
If $\nu(\cdot)$ is chosen from Category 3,
\begin{align}
\left.\frac{E_{b,U}}{N_0}\right|_{C_{fL} = 0} &= \frac{m}{m-1}
\frac{a}{E_w\{\log_2(1 + f(a)|w|^2)\}}.
\end{align}
\end{prop:flashbitenergy}

\vspace{.5cm} \emph{Proof}: We first note that by Jensen's
inequality,
\begin{align}
\frac{C_{fL}(\tsnr, \nu)}{\tsnr} &\le \frac{m\!-\!\!1}{m}
\frac{\nu(\tsnr)}{\tsnr}\log \left( 1 + f\left(
\frac{\tsnr}{\nu(\tsnr)}\right)\right)
\label{eq:cappercostupperbound}
\\
&\stackrel{\text{def}}{=} \zeta(\tsnr, \nu).
\label{eq:cappercostupperbounddef}
\end{align}
First, we consider category 1. In this case, as $\tsnr \to 0$,
$\frac{\tsnr}{\nu(\tsnr)} \to 0$. As shown before, the logarithm in
(\ref{eq:cappercostupperbound}) scales as
$\frac{\tsnr^2}{\nu(\tsnr^2)}$ as $\tsnr \to 0$, and hence
$\zeta(\tsnr,\nu)$ scales as $\frac{\tsnr}{\nu(\tsnr)}$ leading to
\begin{gather}
\lim_{\tsnr \to 0} \frac{C_{fL}(\tsnr,\nu)}{\tsnr} \le \lim_{\tsnr
\to 0} \zeta(\tsnr, \nu) = 0.
\end{gather}
In category 2, $\frac{\tsnr}{\nu(\tsnr)}$ grows to infinity as
$\tsnr \to 0$. Since the $\log(\cdot)$ function on the right hand
side of (\ref{eq:cappercostupperbound}) increases only
logarithmically as $\frac{\tsnr}{\nu(\tsnr)} \to \infty$, we can
easily verify that
\begin{gather}
\lim_{\tsnr \to 0} \frac{C_{fL}(\tsnr,\nu)}{\tsnr} \le \lim_{\tsnr
\to 0} \zeta(\tsnr, \nu) = 0.
\end{gather}
In category 3, $\nu(\tsnr)$ decreases at the same rate as $\tsnr$.
In this case, we have
\begin{align}
\lim_{\tsnr \to 0} \!\!\!\!\frac{C_{fL}(\tsnr,\nu)}{\tsnr} &=
\lim_{n \to \infty}
\frac{C_{fL}\left(\frac{1}{n},\nu\right)}{\frac{1}{n}}
\\
&=\frac{\frac{m-1}{m}E_w\{\lim_{n \to \infty} \log\left(1 +
f(\frac{1}{nv})|w|^2 \right)\}}{a} \label{eq:categ12}
\\
&=\frac{\frac{m-1}{m}E_w\{\log\left(1 + f(a)|w|^2 \right)\}}{a}
\label{eq:categ13}
\end{align}
(\ref{eq:categ12}) is justified
by invoking the Dominated Convergence Theorem and noting the
integrable upper bound
\begin{align}
\left|\log\left(1+f\left(\frac{1}{n\nu}\right)|w|^2\right)\right|
\le 3 \frac{1}{n\nu} |w|^2 \le \frac{3}{\nu} |w|^2 \text{ for } n
\ge 1. \nonumber
\end{align}
The above upper bound is given in the proof of Theorem
\ref{prop:asympcap}. Finally, (\ref{eq:categ13}) follows from the
continuity of the logarithm. \hfill $\square$

Theorem \ref{prop:flashbitenergy} shows that if the rate of the
decrease of the duty cycle  is faster or slower than $\tsnr$ as
$\tsnr \to 0$, the bit energy requirement still increases without
bound in the low-$\tsnr$ regime. This observation is tightly linked
to the fact that the capacity curve $C_L$ has a zero slope as both
$\tsnr \to 0$ and $\tsnr \to \infty$. For improved performance in
the low-$\tsnr$ regime, it is required that the duty cycle scale as
$\tsnr$. A particularly interesting choice is
\vspace{-.2cm}$$\nu(\tsnr) = \frac{1}{a^*} \tsnr$$ where $a^*$ is
equal to the $\tsnr$ level at which the minimum bit energy is
achieved in a non-flashy transmission scheme.
\begin{figure}
\begin{center}
\includegraphics[width = \ffigsize\textwidth]{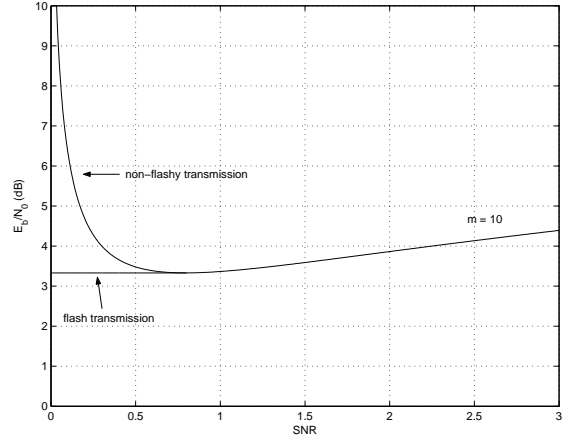}
\caption{Energy per bit $E_{b,U}/N_0$ vs. $\tsnr$ with flash
transmission} \label{fig:flashbitenergy}
\end{center}
\end{figure}
Fig. \ref{fig:flashbitenergy} plots the normalized bit energy
$\frac{E_{b,U}}{N_0}$ as a function of $\tsnr$ for block size $m =
10$. The minimum bit energy is achieved at $\tsnr= 0.8$. For $\tsnr
< 0.8$, flash transmission is employed with $\nu(\tsnr) = 1/0.8 \,
\tsnr$. As observed in the figure, the minimum bit energy level can
be maintained for lower values of $\tsnr$ at the cost of increased
peak-to-average power ratio.

\subsection{Peak Power Constraint on the Pilot}

Heretofore, we have assumed that there are no peak power constraints
imposed on either the data or pilot symbols. Recall that the power
of the pilot symbol is given by
\begin{align}
|x_t|^2 &= \delta^*mP = \sqrt{\xi (\xi + mP)}-\xi
\label{eq:pilotpower}
\end{align}
where $\xi = \frac{m\gamma^2P + (m-1)N_0}{(m-2)\gamma^2}$. We
immediately observe from (\ref{eq:pilotpower}) that the pilot power
increases at least as $\sqrt{m}$ as $m$ increases. For large block
sizes, such an increase in the pilot power may be prohibitive in
practical systems. Therefore, it is of interest to impose a peak
power constraint on the pilot in the following form: \vspace{-.2cm}
\begin{gather}
|x_t|^2 \le \kappa P.
\end{gather}
Since the average power is uniformly distributed over the data
symbols, the average power of a data symbol is proportional to $P$
and is at most $(1-\delta^*)2P$ for any block size. Therefore,
$\kappa$ can be seen as a limitation on the peak-to-average power
ratio. Note that we will allow Gaussian signaling for data
transmission. Hence, there are no hard peak power limitations on
data signals. This approach will enable us to work with a
closed-form capacity expression. Although Gaussian signals can
theoretically assume large values, the probability of such values is
decreasing exponentially.

If the optimal power allocated to a single pilot exceeds $\kappa P$,
i.e., $\delta^*mP > \kappa P \Rightarrow \delta^*m > \kappa$, the
peak power constraint on the pilot becomes active. In this case,
more than just a single pilot may be needed for optimal performance.

In this section, we address the optimization of the number of pilot
symbols when each pilot symbol has fixed power $|x_{t,i}|^2 =\kappa
P \,\, \forall i$. If the number of pilot symbols is $l < m$, then
$\|\x_t\|^2 = l\kappa P$ and, as we know from Section
\ref{sec:trainingscheme},
\begin{align}
\he \sim \CN \left( 0, \frac{\gamma^4 l\kappa P}{\gamma^2 l \kappa
P+ N_0}\right) \text{ and } \hn \sim \CN \left( 0, \frac{\gamma^2
N_0 }{\gamma^2 l \kappa P + N_0}\right). \nonumber
\end{align}
Similarly as before, when the estimate error is assumed to be
another source of additive noise and overall additive noise is
assumed to be Gaussian, the input-output mutual information achieved
by Gaussian signaling is given by
\begin{align}
I_{L,p} = \frac{m-l}{m}E_w\left\{ \log\left( 1 +
g(\tsnr,l)|w|^2\right)\right\}
\end{align}
where $w \sim \CN(0,1)$ and
\begin{gather}
g(\tsnr, l) = \frac{l\kappa(m - l \kappa) \tsnr^2}{(m - l\kappa +
(m-l)l\kappa)\tsnr + m-l}.
\end{gather}
\begin{figure}
\begin{center}
\includegraphics[width = \figsize\textwidth]{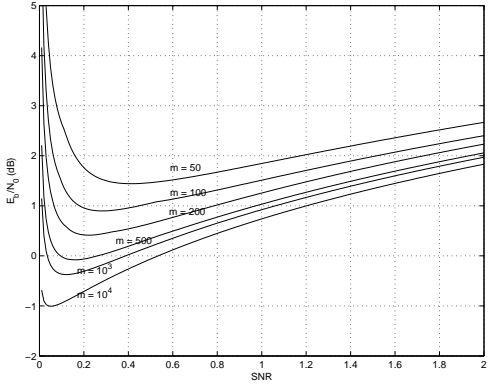}
\caption{Energy per bit $E_{b,U}/N_0$ vs. $\tsnr$ for block sizes of
$m = 50, 100, 200, 10^3, 10^4$. The pilot peak power constraint is
$|x_t|^2 \le 10 P$.} \label{fig:ptrainingbitenergy}
\end{center}
\end{figure}
\begin{figure}
\begin{center}
\includegraphics[width = \figsize\textwidth]{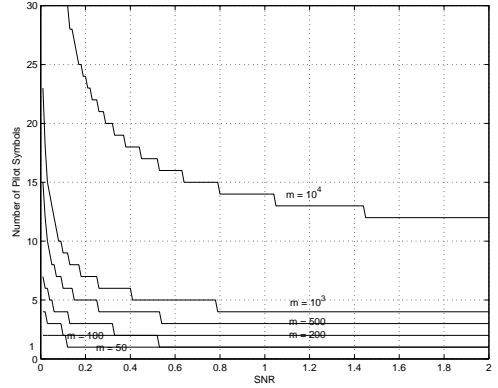}
\caption{Number of pilot symbols per block  vs. $\tsnr$ }
\label{fig:noofpilots}
\end{center}
\end{figure}
The optimal value of the training duration $l$ that maximizes
$I_{L,p}$ can be obtained through numerical optimization. Fig.
\ref{fig:ptrainingbitenergy} plots the normalized bit energy values
$\frac{\tsnr \log2}{I_{L,p}}$ in dB obtained with optimal training
duration for different block lengths. The peak power constraint
imposed on a pilot symbol is $|x_{t}|^2 \le 10 P$. Fig.
\ref{fig:noofpilots} gives the optimal number of pilot symbols per
block. From Fig. \ref{fig:ptrainingbitenergy}, we observe that the
minimum bit energy, which is again achieved at a nonzero value of
the $\tsnr$, decreases with increasing block length and approaches
to the fundamental limit of $-1.59$dB. We note from Fig.
\ref{fig:noofpilots} that the number pilot symbols per block
increases as the block length increases or as $\tsnr$ decreases to
zero.
When there are no peak constraints, $\delta^*m \to m/2$ as $\tsnr
\to 0$. Hence, we need to allocate approximately half of the
available total power $mP$ to the single pilot signal in the
low-power regime, increasing the peak-to-average power ratio. In the
limited peak power case, this requirement is translated to the
requirement of more pilot symbols per block at low $\tsnr$ values.

\begin{table}
\caption{} \label{table:minbitenergy}
\begin{center}\begin{tabular}{|c|c|c|c|p{2.1cm}|} \hline
&  $\frac{E_{b,U}}{N_0}_\text{min}$(dB) & \# of pilots & $\tsnr$ & $\frac{E_{b,U}}{N_0}_\text{min}$ (dB) (\scriptsize{no peak constraints})\\
\hline $m = 50$ & 1.441 & 1 & 0. 41& 1.440
\\ \hline
$m = 100$ & 0.897 & 2 & 0.28& 0.871
\\ \hline
$m = 200$ & 0.413 & 3 & 0.22& 0.404
\\ \hline
$m = 500$ & -0.079 & 5 & 0.16& - 0.085
\\ \hline
$m = 10^3$ & -0.375 & 9 & 0.12& -0.378
\\ \hline
$m = 10^4$ & -1.007 & 44 & 0.05 & -1.008
\\ \hline
\end{tabular}
\end{center}
\end{table}

Table \ref{table:minbitenergy} lists, for different values $m$, the
minimum bit energy values, the required number of pilot symbols at
this level, and the $\tsnr$ at which minimum bit energy is achieved.
It is again assumed that $\kappa = 10$. The last column of the table
provides the minimum bit energy attained when there are no peak
power constraints on the pilot signal. As the block size increases,
the minimum bit energy is achieved at a lower $\tsnr$ value while a
longer training duration is required. Furthermore, comparison with
the last column indicates that the loss in minimum bit energy
incurred by the presence of peak power constraints is negligible.

%
%


\end{document}